\documentstyle[12pt,epsf]{article}

\advance\voffset by -1.5cm
\advance\hoffset by -2.1cm
\def\be{\begin{equation}}
\def\ee{\end{equation}}

\def\Zop{\bbbz}

\def\pmb#1{\setbox0=\hbox{#1}
 \kern-.025em\copy0\kern-\wd0
 \kern.05em\copy0\kern-\wd0
 \kern-.025em\raise.0433em\box0 }

\def\3{\ss}
\def\sq{\hbox{\rlap{$\sqcap$}$\sqcup$}}
\def\qed{\ifmmode\sq\else{\unskip\nobreak\hfil
\penalty50\hskip1em\null\nobreak\hfil\sq
\parfillskip=0pt\finalhyphendemerits=0\endgraf}\fi}
\def\half {\frac{1}{2}}

\def\bbbz {{\sf Z\!\!Z}}

\newcommand{\ket}[1]{|#1\rangle}
\newcommand{\bra}[1]{\langle#1|}
\def\Tr{{\rm Tr}}

\def\ss{\bf S}

\def\drawbox#1#2{\hrule height#2pt 
        \hbox{\vrule width#2pt height#1pt \kern#1pt 
              \vrule width#2pt}
              \hrule height#2pt}

\def\Fund#1#2{\vcenter{\vbox{\drawbox{#1}{#2}}}}
\def\Asym#1#2{\vcenter{\vbox{\drawbox{#1}{#2}
              \kern-#2pt       % line up boxes
              \drawbox{#1}{#2}}}}

\def\funda{\Fund{6.5}{0.4}}
\def\asymm{\Asym{6.5}{0.4}}

\def\symm{\funda\kern-0.4pt\funda}

\begin{document}

\thispagestyle{empty}
\def\thefootnote{\fnsymbol{footnote}}
\begin{flushright}
  hep-th/9906055\\
  CALT-68-2228 \\
  DAMTP-1999-74
 \end{flushright}
\vskip 0.5cm

\begin{center}\LARGE
{\bf Dualities of Type 0 Strings}
\end{center}
\vskip 1.0cm
\begin{center}
{\large  Oren Bergman\footnote{E-mail  address: 
{\tt bergman@theory.caltech.edu}}}

\vskip 0.5 cm
{\it California Institute of Technology\\
Pasadena, CA 91125, USA}

\vskip 1.0 cm
{\large  Matthias R. Gaberdiel\footnote{E-mail  address: 
{\tt M.R.Gaberdiel@damtp.cam.ac.uk}}}

\vskip 0.5 cm
{\it Department of Applied Mathematics and Theoretical Physics \\
University of Cambridge, Silver Street, \\
Cambridge CB3 9EW, U.K.}
\end{center}

\vskip 0.8cm

\begin{center}
June 1999
\end{center}

\vskip 0.8cm

\begin{abstract}
It is conjectured that the two closed bosonic string theories, Type 0A
and Type 0B, correspond to certain supersymmetry breaking orbifold
compactifications of M-theory. Various implications of this
conjecture are discussed, in particular the behaviour of the tachyon
at strong coupling and the existence of non-perturbative fermionic
states in Type 0A. The latter are shown to correspond to bound states
of Type 0A D-particles, thus providing further evidence for the
conjecture. We also give a comprehensive description of the various
Type 0 closed and open string theories.  
\end{abstract}

\vskip 0.8cm 
\begin{center}
\end{center}

\vfill
\setcounter{footnote}{0}
\def\thefootnote{\arabic{footnote}}
\newpage

\renewcommand{\theequation}{\thesection.\arabic
{equation}}

\section{Introduction}
\setcounter{equation}{0} 

Closed string theories are highly constrained by their intrinsic
consistency conditions, in particular modular invariance, and only
finitely many uncompactified theories are known to exist.  Other than
the four supersymmetric theories, there exist nine non-supersymmetric
closed string theories in ten dimensions\footnote{We shall not discuss
the $D=26$ bosonic string in this paper.}: the bosonic Type 0A and
Type 0B theories, as well as seven non-supersymmetric heterotic string
theories \cite{DixHar}. The first two are obtained by projecting Type
IIA and Type IIB, respectively, by the discrete symmetry generated by
the operator $(-1)^{F^s}$, where $F^s$ denotes the spacetime fermion
number; this acts as $\pm 1$ on spacetime bosons and fermions,
respectively, and is therefore equivalent to a rotation by $2\pi$. The
non-supersymmetric heterotic theories correspond to projections of the
two supersymmetric heterotic theories by $(-1)^{F^s}\cdot Y$, where
$Y$ is an automorphism of the internal weight lattice. The resulting
theories are uniquely characterised by their gauge groups, which are
$SO(32)$, $E_8\times SO(16)$, $SO(24)\times SO(8)$, 
$(E_7\times SU(2))^2$, $U(16)$, $E_8$, and $SO(16)\times SO(16)$. All
of the above theories, except for the $SO(16)\times SO(16)$ string,
have tachyons. This has usually been taken to imply either a
perturbative inconsistency, or an instability of the given background.
The latter requires higher order terms in the tachyon potential to
stabilize its vacuum expectation value.

Type 0B string theory, in particular, has attracted recent attention
as a non-supersymmetric setting for a gauge theory -- (super)gravity
correspondence \cite{KT,nonsusy}. This theory possesses Dirichlet
3-branes, whose world-volume field theory is a non-supersymmetric 
four-dimensional gauge theory. In the limit of a large number of 
D3-branes, a variant of the Maldacena conjecture \cite{Maldacena}
suggests a correspondence between the gauge theory on the brane and
Type 0B gravity in a certain background. This has led to novel
insights into the nature of these non-supersymmetric gauge theories. 
An interesting feature of the limit is that the tachyon, due to its
coupling to the R-R fields, actually becomes massive. 

In contrast to closed string theories, the consistency conditions
for open string theories are not completely understood.
The only condition that is normally imposed is
the absence of massless tadpoles in the closed string channel of the
annulus amplitude \cite{PolCai}. This has led, in the supersymmetric
case, to a unique open string theory with gauge group $SO(32)$, namely
Type I. In the non-supersymmetric case there exist a large number of
tadpole-free theories, which can be obtained as world-sheet parity
(orientifold) projections of Type 0A and Type 0B
\cite{BiaSag,BG1,Angel,BFL}; these will be reviewed briefly in section
2.3, where we explain their microscopic realisation in terms of
D-branes. It is not known, however, whether these are the only open
string theories in ten dimensions. Like their closed string ancestors,
these theories contain D-branes, and may therefore be relevant 
to the study of non-supersymmetric gauge theories \cite{BFL}.
\smallskip

There exists by now a convincing picture of a web of string dualities
that connect the various supersymmetric string theories (for a review
see for example \cite{Sen0}). All evidence points towards an
underlying eleven-dimensional structure, known as M-theory, from which
the five ten-dimensional superstring theories emerge in certain
limits. In particular, Type IIA string theory and Heterotic string
theory with gauge group $E_8\times E_8$ correspond to compactification
on a circle and an interval, respectively, whereas Type IIB and both
$SO(32)$ string theories correspond to the zero volume limit of
compactification on a torus and a cylinder, respectively.
It is natural to ask whether the 
non-supersymmetric strings are also part of the web of dualities,
and whether they can be described as certain limits of M-theory as well. 
Early studies indicate that the different non-supersymmetric  string
theories may indeed be related by dualities \cite{BG1,BD}, but the
question of tachyons has so far not been addressed.  

In this paper we argue that Type 0A and Type 0B string theories
can indeed be described as certain supersymmetry breaking
compactifications of M-theory. More specifically, Type 0A corresponds 
to M-theory on ${\bf S}^1/(-1)^{F^s}\cdot S$, where $S$ denotes a
half-shift along the circle, and Type 0B corresponds to the zero
volume limit of M-theory on ${\bf T}^2/(-1)^{F^s}\cdot S$, where the
shift $S$ is along one of the cycles of the torus. In analogy with
Type IIA and Type IIB, the string coupling constant is given by the
radius of the circle in the first case, and by the imaginary part of
the complex structure of the torus in the second case. As in the case
of Type IIB, this realisation of Type 0B suggests that the theory
possesses an $SL(2,\Zop)$ symmetry.

If true, our conjecture implies that the Type 0A tachyon becomes  
massive at strong coupling, and that the theory flows dynamically to
infinite coupling, {\it i.e.} that the eleventh dimension
decompactifies completely. Similarly, the tachyon of Type 0B in nine
dimensions becomes massive for a sufficiently large value of
$g^{(0B)}/R^{(0B)}$. At large coupling and fixed radius the theory
flows then to weakly coupled ten-dimensional Type IIA, whereas at
fixed coupling and small radius it flows to eleven-dimensional
M-theory.  

The relation to M-theory also implies that the Type 0 theories must 
contain non-perturbative fermionic states, even though their
perturbative spectrum is purely bosonic. An independent confirmation
of the existence of such states therefore serves as evidence for the 
conjecture. We shall argue below that such states do indeed exist in
Type 0A, and that they correspond to certain bound states of
D-particles. 

Finally, some of the open string models obtained as world-sheet parity
projections of Type 0A and Type 0B are actually ruled out as a
consequence of the relation to M-theory. For example, the orientifold
of Type 0A is not well-defined at finite coupling, since world-sheet
parity cannot be lifted to a symmetry of the M-theory orbifold. 
\smallskip

The paper is organised as follows. In section~2, we describe the
main features of Type 0 strings, including the perturbative spectrum,
D-branes, and the open string theories that are obtained by
world-sheet parity projection. In section~3 we motivate and state our
conjectures for Type 0A  and Type 0B, and discuss their implications
for the tachyons. In section~4 we describe the non-perturbative
fermionic states of Type 0A. In section~5 we discuss the implications
for the open string models, and in section~6 we present our
conclusions and open questions.

\section{Type 0 Strings}
\setcounter{equation}{0} 

Let us begin by describing the features of Type 0A and Type 0B in
some detail. We shall work in the NS-R formalism, as this is more
appropriate for theories without spacetime supersymmetry.

\subsection{Perturbative spectrum}

The spectra of the Type II theories is given by
\be
\begin{array}{ll}
 {\bf IIA}: & (NS+,NS+)\oplus (R+,R-)\oplus (NS+,R-)\oplus (R+,NS+)
       \\[5pt]
 {\bf IIB}: & (NS+,NS+)\oplus (R+,R+)\oplus (NS+,R+)\oplus (R+,NS+)\,,
\end{array}
\label{typeIIspec}
\ee
where the signs refer to the eigenvalues of $(-1)^{F_L}$ and
$(-1)^{F_R}$, respectively. The effect of $(-1)^{F^s}$ in the
untwisted sector is to retain the bosons ({\it i.e.} the states in
the NS-NS and R-R sectors) and to remove the fermions ({\it i.e.}
the states in the NS-R and R-NS sectors). In the two remaining
sectors, the GSO projection acts in the usual way
\be
\label{gsou}
\begin{array}{ll}
\hbox{NS-NS:} & P_{GSO,U}={1 \over 4} \left( 1 + (-1)^{F_L} \right) 
\left( 1 + (-1)^{F_R} \right) \\
\hbox{R-R:} & P_{GSO,U}={1 \over 4} \left( 1 + (-1)^{F_L} \right) 
\left( 1 \pm (-1)^{F_R} \right) \,,
\end{array}
\ee
where the $+$ sign corresponds to Type IIB, and the $-$ sign to Type 
IIA. In the twisted sector, the effect of $(-1)^{F^s}$ is to reverse
the GSO projection for both left and right-moving sectors. In addition
only the states invariant under $(-1)^{F^s}$ ({\it i.e.} the bosons)
are retained. Thus the states in the twisted sector are again in the
NS-NS and the R-R sector, but their GSO projection is now  
\be
\label{gsot}
\begin{array}{ll}
\hbox{NS-NS:} & P_{GSO,T}={1 \over 4} \left( 1 - (-1)^{F_L} \right) 
\left( 1 - (-1)^{F_R} \right) \\
\hbox{R-R:} & P_{GSO,T}={1 \over 4} \left( 1 - (-1)^{F_L} \right) 
\left( 1 \mp (-1)^{F_R} \right) \,,
\end{array}
\ee
where now the $-$ sign corresponds to Type IIB, and the $+$ sign to
Type IIA. Taking (\ref{gsou}) and ({\ref{gsot}) together, we can
describe the spectrum of Type 0A and Type 0B more compactly as the
subspaces of the NS-NS and R-R sectors that are invariant under the 
GSO-projection 
\be
\label{gso}
\begin{array}{ll}
\hbox{NS-NS:} & P_{GSO}={1 \over 2} \left( 1 + (-1)^{F_L+F_R} \right) 
\\ 
\hbox{R-R:} & P_{GSO}={1 \over 2} \left( 1 \pm (-1)^{F_L+F_R} \right) \,.
\end{array}
\ee
The resulting spectrum is given by
\be
\begin{array}{ll}
 {\bf 0A}: & (NS+,NS+)\oplus (NS-,NS-)\oplus (R+,R-)\oplus (R-,R+)
       \\[5pt]
 {\bf 0B}: & (NS+,NS+)\oplus (NS-,NS-)\oplus (R+,R+)\oplus (R-,R-)\,.
\end{array}
\label{type0spec}
\ee
The NS-NS sector is
the same for the two theories. In particular, the low lying states
consist of the ground state tachyon (that is invariant under
(\ref{gso}) since it is invariant under (\ref{gsot})), and the bosonic
part of the supergravity multiplet, {\it i.e.} the graviton,
Kalb-Ramond 2-form, and dilaton. On the other hand, the R-R sector 
is different for the two theories (as is familiar from Type IIA and
Type IIB). There are no tachyonic states, and the massless states
transform as 
\be
\begin{array}{ll}
 {\bf 0A}: \hspace{2cm} & 
({\bf 8_s} \otimes {\bf 8_c}) \oplus ({\bf 8_c} \otimes {\bf 8_s})
= 2 \cdot {\bf 8_v} + 2 \cdot {\bf 56}\,, \\[5pt]
{\bf 0B}: &
({\bf 8_s} \otimes {\bf 8_s}) \oplus ({\bf 8_c} \otimes {\bf 8_c})
= 2 \cdot {\bf 1} + 2 \cdot {\bf 28} +  {\bf 70}\,.
\end{array}
\ee
In the case of Type 0A, the theory has two 1-forms and two 3-forms in
the R-R sector, whereas Type 0B has two scalars, two 2-forms, and a
4-form (with an unrestricted 5-form field strength). The states in
the R-R sector of Type 0A and Type 0B are therefore doubled 
compared to those in Type IIA and Type IIB.

\subsection{D-branes}

The doubling of the R-R tensor fields suggests that the D-brane
spectrum of Type 0A and Type 0B is doubled compared to that of the
corresponding Type II theories. This is indeed confirmed by a careful
analysis of the boundary states in the two theories \cite{BG1}. Let us
briefly summarise and clarify this analysis.

A typical boundary state in a world-sheet supersymmetric closed
string theory is labelled by an integer $p\in\{-1,\ldots, 9\}$ that
determines the number of Dirichlet and Neumann conditions, and two
spin structures, $\eta$ and $\epsilon$  \cite{PolCai}, where
$\eta$ is either NS-NS or R-R and describes the closed string sector
to which the state belongs, and $\epsilon=\pm$. The defining equations
are\footnote{There are also equations for ghost and superghost 
oscillators, which are independent of $p$ and the spin structures.}
\be
\begin{array}{rl}
 \left.
 \begin{array}{rcl}
  (\alpha^\mu_n - \widetilde{\alpha}^\mu_{-n})\ket{Bp,\epsilon}_\eta
   &=& 0\\[3pt]
  (\psi^\mu_r - i\epsilon\widetilde{\psi}^\mu_{-r})
     \ket{Bp,\epsilon}_\eta &=& 0
 \end{array}
 \right\} & \mu = 0,\ldots,p \\[15pt]
 \left.
 \begin{array}{rcl}
  (\alpha^\mu_n + \widetilde{\alpha}^\mu_{-n})\ket{Bp,\epsilon}_\eta
   &=& 0\\[3pt]
  (\psi^\mu_r + i\epsilon\widetilde{\psi}^\mu_{-r})
     \ket{Bp,\epsilon}_\eta &=& 0
 \end{array}
 \right\} & \mu = p+1,\ldots,9\,.
\end{array}
\ee
Here $n\in\bbbz$, and $r\in\bbbz$ ($\bbbz+1/2$) for $\eta=$R-R (NS-NS).
Up to normalisation, these
equations determine a unique state for each $\eta$ and $\epsilon$, 
\be
 \ket{Bp,+}_{\hbox{\tiny{NS-NS}}}\,,\quad\quad 
 \ket{Bp,-}_{\hbox{\tiny{NS-NS}}}\,,\quad \quad
 \ket{Bp,+}_{\hbox{\tiny{R-R}}}\,,\quad \quad 
 \ket{Bp,-}_{\hbox{\tiny{R-R}}}\,.
\label{bstates}
\ee
The boundary states must be physical closed string states, and thus
must be invariant under the gauge symmetries of the closed string
theory. These include the GSO transformation(s), as well as
orientifold and orbifold transformations, where appropriate. For the
Type II theories this implies that only the combinations 
\be
\begin{array}{ll}
 \ket{Bp,+}_{\hbox{\tiny{NS-NS}}} 
      + \ket{Bp,-}_{\hbox{\tiny{NS-NS}}} \quad & 
    \mbox{for all}\; p \nonumber\\
 \ket{Bp,+}_{\hbox{\tiny{R-R}}} 
      + \ket{Bp,-}_{\hbox{\tiny{R-R}}} \quad & 
    \mbox{for} \; p = \left\{
 \begin{array}{cl}
  \mbox{even} & {\bf IIA} \\
  \mbox{odd} & {\bf IIB}
 \end{array}\right.
\end{array}
\ee
are allowed.\footnote{The normalisation of the state
$\ket{Bp,-}_{\mbox{\tiny{NS-NS}}}$ has been chosen to be minus that
used in \cite{PolCai,BG1}.} In the Type 0 theories on the other hand,
all four boundary states in (\ref{bstates}) are invariant under the
GSO projection (\ref{gso}), with an analogous restriction on the value
of $p$ in the R-R sector.   

The key property of boundary states is that the tree-level amplitude
that describes the propagation from one of them to another can be
reexpressed, through world-sheet duality, as a one-loop open string
vacuum amplitude.  The relevant amplitudes are given by 
\be
\begin{array}{lccl}
 \int dl\, \bra{Bp,\epsilon}\, e^{-lH_{closed}}\,
   \ket{Bp,\epsilon}_{\mbox{\tiny{NS-NS}}} 
   & = & & \half \int {dt\over t}\, \mbox{Tr}_{NS} \Big[e^{-tH_{open}}\Big]
   \\[5pt] 
 \int dl\, \bra{Bp,\epsilon}\, e^{-lH_{closed}}\,
   \ket{Bp,-\epsilon}_{\mbox{\tiny{NS-NS}}} 
   & = & - & \half \int {dt\over t}\, \mbox{Tr}_{R} \Big[e^{-tH_{open}}\Big]
   \\[5pt] 
 \int dl\, \bra{Bp,\epsilon}\, e^{-lH_{closed}}\,
   \ket{Bp,\epsilon}_{\mbox{\tiny{R-R}}} 
   & = & & \half \int {dt\over t}\, \mbox{Tr}_{NS}
   \Big[e^{-tH_{open}}(-1)^F\Big]  
   \\[5pt]
 \int dl\, \bra{Bp,\epsilon}\, e^{-lH_{closed}}\,
   \ket{Bp,-\epsilon}_{\mbox{\tiny{R-R}}} 
   & = & - & \half \int {dt\over t}\, \mbox{Tr}_{R}
   \Big[e^{-tH_{open}}(-1)^F\Big] = 0
  \,.
\end{array}
\label{closedopen}
\ee
This leads to additional consistency conditions, since the 
{\em spectrum} of the open strings that are introduced by the presence of
the boundary states, as well as their {\em interactions} with the
closed strings, must be consistent. Typically, the D-brane
states are therefore linear combinations of the boundary states; these
have to satisfy the following two conditions:
\begin{enumerate}
\item[({\bf a})] For any pair of D-branes, the open string amplitude
 corresponds to a partition function of an open string theory, 
 {\it i.e.} it takes the form of a sum of traces over sets 
 of open string states of the time evolution operator
 $e^{-tH_{open}}$.
\item[({\bf b})] The open-closed vertex that describes the
 joining of the two ends of an open string to form a closed string
 is well-defined on physical states.
\end{enumerate}
One is usually also interested in D-branes that are {\em stable}; 
this is equivalent to the condition that the spectrum of open 
strings that begin and end on the {\em same} D-brane is free of
tachyons. If the underlying theory is supersymmetric, one may also want
to impose the condition that the D-branes preserve some of the
supersymmetry, {\it i.e.} that they are {\em BPS
saturated}; this requires that the spectrum of open strings beginning
and ending on the D-brane is supersymmetric.

For the Type II theories the BPS and stability conditions are
equivalent.\footnote{In other theories, such as Type I, and certain
orbifolds of Type II, a tachyon-free open string spectrum can also be
achieved without supersymmetry, resulting in stable non-BPS
D-branes \cite{SenAll,BG2,WittenK,FGLS,SenRev}. } These select
the linear combinations 
\be
 {\bf Type}\;{\bf II}:\quad
 \ket{Dp} = \Big(\ket{Bp,+}_{\hbox{\tiny{NS-NS}}} 
             + \ket{Bp,-}_{\hbox{\tiny{NS-NS}}}\Big) \pm
            \Big(\ket{Bp,+}_{\hbox{\tiny{R-R}}} 
             + \ket{Bp,-}_{\hbox{\tiny{R-R}}}\Big) \,,
\label{DII}
\ee
where $p$ is even (odd) for Type IIA (IIB). The relative sign of the
R-R and NS-NS components determines the sign of the R-R charge carried
by the D-brane, and therefore differentiates a brane from an
anti-brane.  By applying (\ref{closedopen}), it immediately follows
that ({\bf a}) is satisfied. It is also generally believed that 
({\bf b}) is satisfied for these states, although this has never been
completely verified to our knowledge.

The NS-NS component of (\ref{DII}) actually satisfies ({\bf a}) (and
presumably ({\bf b})) by itself, but is non-supersymmetric, and in
fact unstable due to the presence of an open string tachyon. The
combinations given by 
\be
 \ket{\widehat{Dp}} = \Big(\ket{Bp,+}_{\hbox{\tiny{NS-NS}}} 
             + \ket{Bp,-}_{\hbox{\tiny{NS-NS}}}\Big) 
\label{unstable}
\ee
therefore correspond to unstable (and non-BPS) D-branes. As such, they
are not really seen as part of the spectrum of Type II string
theory. Nevertheless, unstable D-branes are very useful as
intermediaries in describing descent relations among D-branes in the
Type II theories \cite{SenRev,Horava}. 

For the Type 0 theories, property ({\bf a}) again implies that one
should take either a combination of an NS-NS state and an R-R state,
or just an NS-NS state, but again only the former will be stable. As
for Type II theories, property ({\bf b}) has not been completely 
analysed. However, one immediate implication of ({\bf b}) is that the
spectrum of open strings that begin and end on the same D-brane
must be free of fermions; if this were not so, closed string
fermions, which are absent in the closed string spectrum, would emerge
from the open-closed vertex.  This condition implies (see \cite{BG1}
for a detailed discussion) that of the various (stable) states that are
consistent with ({\bf a}), only eight linear combinations remain 
\be
\label{D0p}
\ket{Bp,\epsilon }_{\hbox{\tiny{NS-NS}}} 
      \pm \ket{Bp,\epsilon' }_{\hbox{\tiny{R-R}}} \,,
\ee
where $p$ is even (odd) in Type 0A (0B).

The number is further reduced by imposing condition ({\bf a}) on
the amplitudes involving two {\em different} D-branes.\footnote{This 
condition was not considered in \cite{BG1};
however all D-branes that were relevant for the description of the
duality in that paper satisfied it. See also \cite{BG2} for a
related discussion in a different context.} This requires that, out
of the eight combinations above, we keep only the four with 
$\epsilon = \epsilon'$, or only the four with 
$\epsilon = -\epsilon'$. This is because the amplitude for
one of the former states to propagate into one of the latter states
corresponds to a trace over open string states with an insertion
of $(-1)^F$, rather than $1$ or a projection operator,
and therefore violates condition ({\bf a}). On the other hand,
amplitudes involving only the states with $\epsilon = \epsilon'$ (or
$\epsilon = - \epsilon'$) will correspond  to traces over projected
sets of open string states, and will therefore satisfy ({\bf a}).
Let us fix our conventions by keeping the four combinations
with $\epsilon = \epsilon'$; these describe the two possible
D-branes (and D-antibranes),
\begin{equation}
\label{D0}
{\bf Type}\;{\bf 0}:\quad
\begin{array}{rcl}
 \ket{Dp,+} &=& \ket{Bp,+}_{\hbox{\tiny{NS-NS}}}
        \pm \ket{Bp,+}_{\hbox{\tiny{R-R}}} \\[5pt]
 \ket{Dp,-} &=& \ket{Bp,-}_{\hbox{\tiny{NS-NS}}}
        \pm \ket{Bp,-}_{\hbox{\tiny{R-R}}} \,.
\end{array}
\end{equation}

The above restriction was also obtained in \cite{KT} using a
different method. There it was argued that D-branes that carry
opposite charges with respect to the R-R fields in the twisted
sector should also couple to the tachyon in the twisted NS-NS sector 
with opposite signs. The signs of these charges are in turn related
to the spin structures as follows; the two R-R boundary states with
$\epsilon' = \pm$ carry equal (unit) charges under the untwisted R-R
fields, and opposite charges under the twisted R-R fields. (This is an
obvious consequence of the fact that Type II D-branes (\ref{DII}) are
charged only under the untwisted fields.) Similarly, the two NS-NS
boundary states couple to the tachyon with opposite signs. The
argument of \cite{KT} then implies that the two spin structures must
be correlated, reducing the allowed D-brane states to those in
(\ref{D0}).   

The open string spectra associated with these D-branes can easily be 
read off from the two relevant amplitudes,
\be
\begin{array}{lccl}
 \int dl\, \bra{Dp,\epsilon}\, e^{-lH_{closed}}\, \ket{Dp,\epsilon}
   & = & & \int {dt\over t}\, \mbox{Tr}_{NS}\Big[ e^{-tH_{open}}
   {1\over 2}(1+(-1)^F) \Big] \\[10pt]
 \int dl\, \bra{Dp,\epsilon}\, e^{-lH_{closed}}\, \ket{Dp,-\epsilon}
   & = & - & \int {dt\over t}\, \mbox{Tr}_{R}\Big[ e^{-tH_{open}}
   {1\over 2}(1+(-1)^F) \Big] \,.
\end{array}
\ee
In particular, open strings between like D-branes are purely bosonic, 
whereas those between unlike D-branes (of equal $p$) are purely
fermionic. The corresponding massless excitations are the appropriate
dimensional reduction of a ten-dimensional vector and a
ten-dimensional Majorana-Weyl spinor, respectively. For
brane-antibrane amplitudes the sign in front of $(-1)^F$ is reversed.
Thus the open string between a brane and an anti-brane of the same
type contains a tachyon (but no massless vectors), and the open string
between a brane of one type and an anti-brane of the other has
fermions of the opposite chirality.

\subsection{Open strings}

The various Dirichlet-branes, and in particular the D9-branes, play a
crucial role in the construction of the (open string) orientifold models.
Although these models are somewhat peripheral to the main line of the 
paper, we shall nevertheless use the occasion to explain how they
can be understood in terms of D-branes. The models fall into four
classes, three of which are obtained from Type 0B, and one from Type
0A. 
\smallskip

\noindent\underline{\bf 0B model 1}:
Like Type IIB string theory, Type 0B is invariant under the
world-sheet parity operator $\Omega$, and we can therefore consider
the corresponding projection. The massless string states that are
invariant under $\Omega$ are the graviton, dilaton and the R-R 2-form
in the untwisted  sector, as well as the tachyon and the R-R 2-form in
the twisted sector. The one-loop vacuum amplitude of the new theory
has a contribution from the Klein bottle,  
\be
 K = \int {dt\over 2t}\Tr_{closed}
   \Big[e^{-tH_{closed}}{1\over 2}(1+(-1)^{F_L+F_R})
        {1\over 2}\Omega\Big] \,.
\label{klein}
\ee
When interpreted in the tree channel, this gives rise to a massless
NS-NS tadpole. The theory does not have a massless R-R tadpole, as
that would arise from an insertion of $((-1)^{F_L} + (-1)^{F_R})$ in
the trace. The presence of the massless NS-NS tadpole does not render
the theory necessarily inconsistent; it only means that the vacuum has
to be shifted by the Fischler-Susskind mechanism \cite{FS}. 
Alternatively, one can introduce 64 D9-branes to cancel the NS-NS
tadpole. However, as the net R-R charge must still vanish, 
32 of them must be anti-D9-branes. Since there are two kinds of
D9-branes in Type 0B, one can introduce $n$ of one kind and $32-n$ of
the other, together with their respective anti-branes. From the
analysis of the previous subsection we can easily determine the
spectrum of open string in this theory. In particular, the gauge group
is $(SO(n)\times SO(32-n))^2$; all the other low lying states are
given in table~1.

For the specific choice $n=0$ (or $n=32$), the resulting theory is
purely bosonic, and has a gauge group $SO(32)\times SO(32)$.  It is
further singled out by the property that one of its D-strings possesses
low-lying modes that reproduce the world-sheet theory of the $D=26$ 
bosonic
string compactified on an $SO(32)$ lattice to ten dimensions. This has
led to the suggestion that the two models might be related by a
strong/weak coupling duality \cite{BG1}.
\smallskip

\noindent\underline{\bf 0B model 2}:
The action of $\Omega$ can also be combined with another $\Zop_2$
symmetry. In particular, we can consider the operator
$\widehat{\Omega}$ that differs from $\Omega$ by a sign in the R-R 
sector \cite{BiaSag}, and can therefore be described as
$\widehat{\Omega}=\Omega (-1)^{F_R^s}$, where $F_R^s$ is the
right-moving part of the space-time fermion number
operator. ($\widehat{\Omega}$ is of order two since
$\widehat{\Omega}^2 = (-1)^{F^s}\equiv 1$ in Type 0B.)  The invariant
massless fields now include the two scalars and the 4-form, instead of
the two 2-forms. The Klein bottle contribution to the vacuum amplitude
is again given by (\ref{klein}), with $\Omega$ replaced by
$\widehat{\Omega}$, and therefore the massless R-R tadpole
vanishes. 
The theory has a NS-NS tadpole, but it is tachyonic rather that
massless. Since the theory has a closed string tachyon anyway, this is
believed to be harmless, and there is therefore no need to introduce
any branes at all. The resulting theory is thus a closed bosonic
string theory whose low-lying spectrum is shown in table~1. 
\smallskip

\noindent\underline{\bf 0B model 3}:
Finally, there exists a third $\Zop_2$ operator $\Omega'$ that
can be described as  $\Omega' = \Omega (-1)^{F_R}$,
\cite{BiaSag,Angel,BFL}. ($\Omega'$ is of order two since 
$(\Omega')^2 = (-1)^{F_L+F_R}\equiv 1$ in Type 0B.)  
Because of the $(-1)^{F_R}$, this projection removes the tachyon.
At the massless level, the surviving fields are the graviton and dilaton
from the NS-NS sector, and a single scalar, a single 2-form, and a 4-form
with anti-self-dual field strength from the R-R sector. Replacing
$\Omega$ by $\Omega'$ in the Klein bottle (\ref{klein}) results in a
trace with $((-1)^{F_L} + (-1)^{F_R})$, and therefore the theory has 
a massless R-R tadpole, but no massless NS-NS tadpole. For
consistency, this tadpole {\em must} be cancelled by the introduction
of D9-branes. Since the operator $(-1)^{F_R}$ maps one kind of
D9-brane ($\ket{D9,+}$ say) to the other ($\ket{D9,-}$) (see
\cite{BG1,BFL}), we are forced to introduce them in pairs. In order to
cancel the R-R tadpole, a net number of $32$ D9-brane pairs is
required. The inclusion of these D9-branes necessarily introduces a
massless NS-NS tadpole which cannot be cancelled (but which, as
before, does not render the theory necessarily inconsistent). 
The resulting gauge group is $U(32)$, and the theory is completely
free of tachyons. More generally, since the total number of D9-branes
is not restricted, one can add  an arbitrary but equal number of
D9-brane pairs and anti-D9-brane pairs to the background, without
introducing a R-R tadpole. The resulting gauge group is then
$U(n)\times U(32+n)$, where $n$ denotes the number of anti-D9-brane
pairs. This theory has tachyons in the bi-fundamental and its complex
conjugate. 
\smallskip

\noindent\underline{\bf 0A model}: Unlike its supersymmetric cousin,
Type 0A is also invariant under $\Omega$, and we can therefore
consider its orientifold \cite{BiaSag}. In the NS-NS sector, $\Omega$
acts in the same way as for Type 0B, and therefore leaves the
graviton, dilaton, and tachyon invariant. In the R-R sector, $\Omega$
exchanges the two 1-forms and the two 3-forms, and therefore only the
symmetric combinations are invariant.  The resulting theory has a
Klein bottle given by (\ref{klein}), and therefore a massless NS-NS
tadpole, but no R-R tadpole. This can be cancelled by the inclusion of
open string states, leading to the gauge group $SO(n)\times SO(32-n)$,
where $n\in\{1,\ldots, 32\}$. The open string spectrum also contains 
tachyons
in the singlet representation, as well as in the symmetric tensor
representations of either $SO$ factor. There are also massless
fermions in the bi-vector representation. In terms of D-branes, these
open strings can be understood to originate from $n$ unstable D9-branes
of one type ($\ket{\widehat{D9},+}$, say) and $32-n$ unstable D9-branes
of the other type ($\ket{\widehat{D9},-}$).  The tachyons arise from
open strings which begin and end on like D9-branes (and in particular
open strings beginning and ending on the same D9-brane), and the
massless fermions arise from open strings which stretch between unlike
D9-branes. These 9-branes only carry NS-NS charge and can
therefore cancel the NS-NS tadpole.

Unlike in the case of Type 0B, the operators 
$\widehat{\Omega}=\Omega (-1)^{F_R^S}$ and $\Omega'= \Omega(-1)^{F_R}$
do not give new open string models when applied to Type 0A. The former
gives precisely the same Type 0A model as before, except that now only
the {\em anti-symmetric} combinations of the 1-forms and 3-forms
survive. The latter operator generates a $\Zop_4$ rather than $\Zop_2$
symmetry, since ${\Omega'}^2 = (-1)^{F_L+F_R}$ is $+1$ in the NS-NS
sector, but $-1$ in the R-R sector. The orbifold by $(-1)^{F_L+F_R}$
gives Type 0B, and the resulting model is therefore one of the Type 0B
open string theories.  

\begin{table}[htb]
\begin{center}
\begin{tabular}{|l|c|c|c|l|} \hline
 string & gauge group & tachyons & fermions & R-R fields \\ \hline
 &&&&\\[-10pt]
 0A & -- & 1 & -- & $2({\bf 8_v} \oplus {\bf 56})$ \\[3pt] 
 0B & -- & 1 & -- & $2({\bf 1} \oplus {\bf 28}) \oplus  {\bf 70}$\\[5pt]
 0A$/\Omega$ & $SO(n)\times SO(32-n)$ & 
  $\left\{
   \begin{array}{c}
   ({\bf 1},{\bf 1}) \\
   (\,\symm \, ,{\bf 1})  \\
   ({\bf 1},\symm \,)
  \end{array}
  \right\}$
  & $(\,\funda \,,\funda \,)$ & ${\bf 8_v} \oplus {\bf 56}$ \\[20pt]
 0B$/\Omega$ & $[SO(n)\times SO(32-n)]^2$ & 
  $\left\{
   \begin{array}{c}
    ({\bf 1},{\bf 1},{\bf 1},{\bf 1}) \\
    (\funda \,,{\bf 1},\funda \,,{\bf 1})\\
    ({\bf 1},\funda \,,{\bf 1},\funda \,) 
   \end{array}
   \right\}$
  & 
  $\left\{
   \begin{array}{c}
   (\,\funda \,,\funda \,,{\bf 1},{\bf 1}) \\
   (\,\funda \,,{\bf 1},{\bf 1},\funda \,) \\
   ({\bf 1},{\bf 1},\funda \,,\funda \,) \\
   ({\bf 1},\funda\,,\funda \,,{\bf 1}) 
   \end{array}
   \right\}$
  & $2({\bf 28})$  \\[34pt]
 0B$/\widehat{\Omega}$ 
   & -- & 1  & -- & $2({\bf 1}) \oplus {\bf 70}$ \\[15pt]
 0B$/\Omega'$ &  $U(n)\times U(32+n)$ & $(\,\funda \,,\overline{\funda}
   \,) \oplus 
    \mbox{c.c}$
  & 
  $\left\{
   \begin{array}{c}  
   (\,\symm \,,{\bf 1}) \oplus \mbox{c.c.} \\
   ({\bf 1},\asymm \,) \oplus \mbox{c.c} \\
   (\,\funda \,,\overline{\funda} \,) \oplus \mbox{c.c.}
  \end{array}
  \right\}$
  & ${\bf 1} \oplus {\bf 28} \oplus {\bf 35_+}$  \\[20pt]\hline
\end{tabular}
\caption{Type 0 string theories -- low-lying spectrum.}
\end{center}
\end{table}

\section{From M-theory to Type 0}
\setcounter{equation}{0}

Given the relation of Type 0A string theory to Type IIA, one may ask
whether the strong coupling dynamics of the former is also
eleven-dimensional in character. The obvious candidate for such an
eleven-dimensional theory is the orbifold of M-theory by
$(-1)^{F^s}$. Type 0A would then correspond to the compactification of
this orbifold on a circle of radius $R^{(M)}$; a simple argument,
however, rules this scenario out.

In the limit $R^{(M)}\rightarrow 0$, the M-theory orbifold should
reproduce the massless spectrum of Type 0A, and therefore in
particular two 1-forms, a 2-form, and two 3-forms. Since the action of
the orbifold is independent of $R^{(M)}$, these can only arise by 
Kaluza-Klein reduction of massless fields in eleven dimensions. The
invariant sector of the original massless eleven-dimensional
supergravity multiplet consists of the graviton and the 3-form, which
upon Kaluza-Klein reduction are only able to produce a single
1-form, a 2-form, and a single 3-form. The additional 1-form and
3-form must come from a `twisted sector' of the orbifold. For the
1-form this requires either a massless 1-form or 2-form in eleven
dimensions, and for the 3-form it requires either a massless 3-form or
4-form.  In either case, this would lead to additional massless states
in ten dimensions that are not present in Type 0A. Therefore Type 0A
cannot correspond to the circle compactification of M$/(-1)^{F^s}$;
the same argument actually implies more generally that Type 0A cannot
be obtained by ordinary Kaluza-Klein dimensional reduction from 
{\em any} theory in eleven 
dimensions.

\subsection{Smooth orbifolds}

It is however possible to obtain insight into the relation between
Type 0A and M-theory by smoothing out the orbifold which produces Type
0A from Type IIA. The general prescription for doing this is to
compactify an extra dimension on a circle of radius $R$, and combine
the action of the original discrete symmetry with that of a half-shift
along the circle. If we denote the original ($\Zop_2$) symmetry by
$g$, the new symmetry is given by $g\cdot S$, where $S$ denotes a
half-shift along the circle. The untwisted sector corresponds to
strings with integer winding number, and consists of $g$-even states
with even Kaluza-Klein momentum, and $g$-odd states with odd
Kaluza-Klein momentum. In addition, there is a twisted sector arising
from $g$-twisted strings of half-odd-integer winding number, in which
the states are again $g$-even with even momentum, and $g$-odd with odd
momentum. In the limit $R\rightarrow 0$, the $g$-odd states of both
the untwisted and the twisted sector decouple, and states with
different winding numbers become degenerate. In this limit we
therefore reproduce the spectrum of the original orbifold by $g$
compactified on a circle of vanishing radius (or, equivalently the
T-dual of this orbifold at infinite radius).

For $g=(-1)^{F^s}$ the resulting nine-dimensional theories obtained
from Type IIA and Type IIB were first considered in \cite{Rohm}, and
are related by a Wick rotation to the finite temperature strings
considered in \cite{AtiWit}. Their spectrum is given by
\be
\begin{array}{rl}
 \mbox{untwisted:} & 
 \begin{array}{lll}
  (NS+,NS+)\oplus  (R+,R\mp)\,, & p_9=2m/R_9^{(II)}\,, & w_9=nR_9^{(II)} 
    \\[3pt]
  (R+,NS+) \oplus  (NS+,R\mp)\,, & p_9=(2m+1)/R_9^{(II)} \,, & 
     w_9=nR_9^{(II)}
 \end{array} \\[20pt]
 \mbox{twisted:} &
 \begin{array}{lll}
  (NS-,NS-)  \oplus  (R-,R\pm)\,, & p_9=2m/R_9^{(II)}\,, & 
      w_9=(n+1/2)R_9^{(II)} \\[3pt]
  (R-,NS-)  \oplus  (NS-,R\pm)\,, & p_9=(2m+1)/R_9^{(II)} \,, & 
      w_9=(n+1/2)R_9^{(II)} 
  \,,
 \end{array} 
\end{array}
\label{smoothspec}
\ee
where the upper (lower) sign corresponds to Type IIA(IIB). In the limit
$R_9^{(II)}\rightarrow\infty$ one regains the spectrum of
ten-dimensional Type IIA(B) (\ref{typeIIspec}), and in the limit
$R_9^{(II)}\rightarrow 0$ one obtains the spectrum of Type 0A(B) 
for $R_9^{(0)}=0$ (\ref{type0spec}). In particular, all the fermions 
are removed, and additional R-R fields become massless in this limit.
In addition, there is a tachyon in the twisted sector when
$R_9^{(II)}<\sqrt{2}$. For $R_9^{(II)}>\sqrt{2}$ these
theories have a finite vacuum energy, which decreases monotonically
with $R_9^{(II)}$, and vanishes in the supersymmetric limit
$R_9^{(II)}\rightarrow\infty$ \cite{Rohm}. In this regime, the
nine-dimensional theories therefore flow back to the ten-dimensional 
Type II theories. When $R_9^{(II)}<\sqrt{2}$, on the other
hand, the vacuum energy diverges due to the presence of the tachyon. 
The theories are therefore either inconsistent in this regime, or they
undergo a phase transition (akin to the Hagedorn transition at finite
temperature), by tachyon condensation, into a stable vacuum.  

\subsection{Type 0A}

Strictly speaking, the above orbifold procedure is reliable only at
weak coupling, for which the corresponding spectrum can be trusted; 
we can therefore identify
\be
  \mbox{{\bf Type IIA} on}\;\; {\bf S}^1/(-1)^{F^s}\cdot S 
     \Big|_{g^{(IIA)}=0,R^{(IIA)}_9=0} 
   = 
  \mbox{{\bf Type 0A} on}\;\; {\bf S}^1 \Big|_{g^{(0A)}=0,R^{(0A)}_9=0} \,,
\label{ident}
\ee
and similarly for Type IIB. On the other hand Type IIA string theory
corresponds to the compactification of M-theory on a circle. From the
point of view of  M-theory, the above nine-dimensional theory thus
corresponds to the compactification on 
${\bf S}^1\times{\bf S}^1/(-1)^{F^s}\cdot S$, where both circles are
small. Let us denote the radius of the ordinary circle (when measured 
with respect to the eleven-dimensional metric) by $R_{10}^{(M)}$, and
the radius of the orbifolded circle by $R_{9}^{(M)}$. The
(ten-dimensional) coupling constant and radius in the
nine-dimensional theory are then given by
\be
   g^{(IIA)} = (M_{11}R_{10}^{(M)})^{3/2} \;,\quad
   R^{(IIA)}_9 = (M_{11}R_{10}^{(M)})^{1/2} R^{(M)}_9 \,,
\label{IIAvars}
\ee
where $M_{11}$ is the eleven-dimensional Planck mass.
By exchanging the roles of the two circles (see Fig.~1(a)), we can
describe this theory equivalently as an ordinary circle
compactification of a ten-dimensional theory. 
The identification (\ref{ident}) for zero coupling and radius 
suggests then that this ten-dimensional theory is Type 0A, where 
the coupling constant and radius are given by\footnote{The factor
of $1/2$ is due to the half-shift $S$.} 
\be
   g^{(0A)} = (M_{11}R_{9}^{(M)}/2)^{3/2} \;,\quad
   R_9^{(0A)} = (M_{11}R_{9}^{(M)}/2)^{1/2} R^{(M)}_{10} \,.
\ee
We are thus led to conjecture that 
\begin{center}
  {\bf Type 0A} = {\bf M-theory} on
   ${\bf S}^1/(-1)^{F^s}\cdot S$ .
\end{center}
\begin{figure}[htb]
\epsfxsize=5 in
\centerline{\epsffile{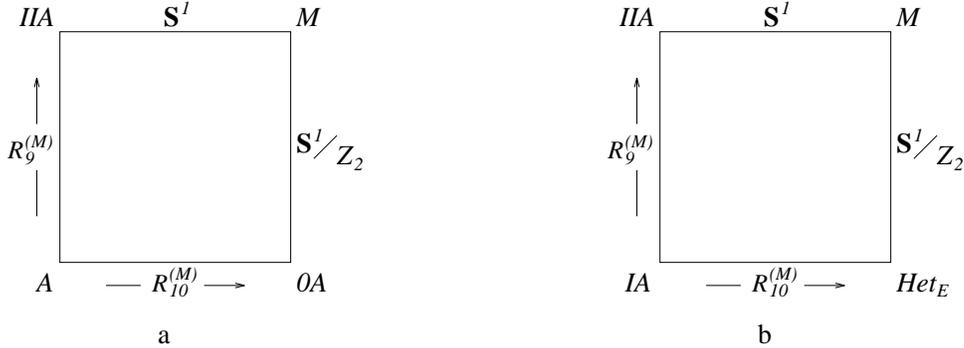}}
\caption{The `9-10 flip' in M-theory; 
(a) Type 0A, (b) Heterotic $E_8\times E_8$.}
\end{figure}
In essence, the conjecture means that Type 0A string theory
corresponds to M-theory compactified on a circle of radius
$R^{(M)}/2$, where we impose periodic boundary conditions for bosons,
and anti-periodic boundary conditions for fermions.\footnote{A 
Euclidean version of Type 0A can then be thought of as M-theory at
finite temperature, where $T\sim 1/R^{(M)}$.} The massless
spectrum of the orbifold contains the bosonic part of the Type IIA
supergravity multiplet, namely a ten-dimensional graviton, a
Kalb-Ramond field and a dilaton, together with a (single) 1-form field
and a 3-form field. The action of the orbifold symmetry does not have
any fixed points, and the resulting theory therefore corresponds to a
closed string theory (if it defines a string theory at all). Since the
orbifold also breaks supersymmetry, this closed string theory can only
be either Type 0A or Type 0B, or one of the non-supersymmetric
heterotic theories. As we have just seen, the massless spectrum
contains the bosonic part of the Type IIA supergravity multiplet, and
therefore the only possibility is Type 0A.

M-theory orbifolds, like string theory orbifolds, usually require
the inclusion of `twisted sector' states for consistency
\cite{HorWit,DasMuk,Wit1}. However, unlike string theory where
twisted sectors arise from strings that close up to the action by an
element of the orbifold group, the origin of twisted sectors in
M-theory orbifolds is much less understood. Instead, the twisted
sectors of M-theory orbifolds are usually determined by indirect
arguments such as the requirement that certain symmetries are
protected from anomalies. (This is actually somewhat analogous to the
way in which twisted sectors of asymmetric string orbifolds are
determined by the requirement of modular invariance.) The paradigm for
this is the compactification of M-theory on ${\bf S}^1/\Zop_2$
\cite{HorWit}, where the non-trivial element of the orbifold group
reflects the compact coordinate and reverses the sign of the 3-form,
\be 
x^{10}\rightarrow - x^{10} \;,\quad C^{(3)} \rightarrow - C^{(3)} \,.
\label{reflect}
\ee
If we compactify the theory on an additional circle to nine
dimensions and exchange the roles of the two compact directions,
similar arguments as above suggest that the M-theory orbifold 
corresponds to the $E_8\times E_8$ heterotic string (see Fig.~1(b)). 
Indeed, the untwisted sector of the M-theory orbifold accounts for the
$N=1$ gravity supermultiplet, and the twisted sectors provide the two 
$E_8$ gauge multiplets. These can also be shown to be necessary in
order to cancel the gravitational anomaly due to the boundaries.
In the present case, however, the orbifold by $(-1)^{F^s}\cdot S$ does
not introduce a gravitational anomaly, and it is therefore not clear 
what symmetry would be violated if the twisted sector were absent.

In the limit $R^{(M)}\rightarrow 0$ the twisted sector of our orbifold
should include a tachyon and an additional massless 1-form and 3-form
field. This is now possible, in principle, since the action of the
orbifold does depend on $R^{(M)}$, and so fields which are massive at
large $R^{(M)}$ may become massless, or even tachyonic, when 
$R^{(M)}\rightarrow 0$. Conversely, the additional R-R fields of Type 
0A should become massive at non-zero coupling, and the tachyon should
become massive at sufficiently strong coupling. Therefore the
tachyonic instability disappears in the strong coupling regime, and
the theory proceeds to flow back to eleven dimensions.

\subsection{T-duality and Type 0B}

It is well known that the ${\bf S}^1$ compactifications of Type IIA  
and Type IIB string theories are related by T-duality; this symmetry 
inverts the radius of the circle, exchanges winding and momentum
states, and changes the sign of $(-1)^{F_R}$ in the right-moving
R-sector. Similarly, T-duality on a circle relates Type 0A and Type
0B to one another.   

The transformation of the nine-dimensional theories described in
subsection 3.1 under T-duality is somewhat more complicated. T-duality
still exchanges winding and momentum states, and this requires that it
maps $R_9^{(II)}\rightarrow 2/R_9^{(II)}$ so that even momentum
becomes integer winding, and odd momentum becomes half-odd-integer
winding. In the T-dual theory all spacetime bosons therefore have
integer winding, whereas all spacetime fermions have half-odd-integer 
winding. Furthermore, states which are even under $(-1)^{F_R}$ have
even momentum, and those odd under $(-1)^{F_R}$ carry odd
momentum. It thus follows that the theory T-dual to Type
IIA (IIB) on ${\bf S}^1/(-1)^{F^s}\cdot S$ is given by 
Type 0B (0A) on ${\bf S}^1/(-1)^{F_R}\cdot S$ (compare also \cite{Green}).

The `9-10 flip' (Fig.~1(a)), which allowed us to identify Type 0A as
M-theory on ${\bf S}^1/(-1)^{F^s}\cdot S$, is equivalent to the
sequence of duality transformations given by $T'\circ S\circ T$
\cite{Polchinski}, where
\be
\begin{array}{rll}
 T: & R\rightarrow 1/R\,, & g\rightarrow g/R \\
 T': & R\rightarrow 2/R\,, & g\rightarrow g/R \\
 S: & R\rightarrow g^{-1/2}R\,, & g\rightarrow 1/g \,.
\end{array}
\ee
It therefore implies an S-duality relation between Type 0B on 
${\bf S}^1$ and Type 0B on ${\bf S}^1/(-1)^{F_R}\cdot S$. 
In the infinite radius limit this reduces to the statement that Type
0B is symmetric under S-duality. 

The T-duality relation between Type 0A and Type 0B suggests that the
tachyon of Type 0B becomes massive when $g^{(0B)}/R^{(0B)}$ is
sufficiently large.\footnote{We are assuming here that T-duality is a
symmetry beyond the perturbative regime.} In this regime, the theory
behaves differently depending on whether $g^{(0B)}\gg 1$ and
$R^{(0B)}$ is fixed, or whether $R^{(0B)}\ll 1$ and $g^{(0B)}$ is 
fixed. In the first limit, the theory is related by S-duality
to Type 0B on ${\bf S}^1/(-1)^{F_R}\cdot S$, with a weak coupling and
small radius given by
\be
 g'^{(0B)} = 1/g^{(0B)} \ll 1 \;, \quad 
 R'^{(0B)} = R^{(0B)}/\sqrt{g^{(0B)}} \ll 1 \,.
\ee
This is in turn related by T-duality to Type IIA on 
${\bf S}^1/(-1)^{F^s}\cdot S$ at weak coupling and large radius given by
\be
 g^{(IIA)} = 1/(\sqrt{g^{(0B)}}R^{(0B)}) \ll 1 \;, \quad
 R^{(IIA)} = 2\sqrt{g^{(0B}}/R^{(0B)} \gg 1 \,.
\ee
Type 0B therefore flows to weakly coupled ten-dimensional Type IIA
in this limit. In the second limit, the theory is related by T-duality
to Type 0A at large radius and strong coupling,
\be
 g^{(0A)} = g^{(0B)}/R^{(0B)} \gg 1 \;,\quad
 R^{(0A)} = 1/R^{(0B)} \gg 1  \,,
\ee
which, as we have already seen, flows to eleven-dimensional M-theory.
\smallskip

In terms of M-theory, again because of T-duality, Type 0B corresponds 
to 
\begin{center}
 {\bf Type 0B} = {\bf M-theory} on
   ${\bf T}^2/(-1)^{F^s}\cdot S \Big|_{V=0}$ \,,
\end{center}
where the shift acts on one of the cycles of the torus, and the
coupling constant is identified with the imaginary part of the complex
structure of the torus. This then suggests that Type 0B 
is symmetric not only under S-duality, but under the full
$SL(2,\bbbz)$ symmetry. 
\smallskip

In Type IIB S-duality exchanges the fundamental string
with the D-string, and likewise the NS-NS 2-form with the R-R
2-form. The spectrum of open strings that begin and end on the
D-string contains 8 massless bosons and 8 massless fermions of either
chirality, and these agree precisely with the low-lying excitations
of the fundamental string.

In Type 0B there are {\em two} D-strings, $\ket{D1,+}$ and
$\ket{D1,-}$, and correspondingly {\em two} R-R 2-forms.
Neither of these D-strings has the correct degrees of freedom
to correspond to a fundamental Type 0B string, as neither possesses
fermionic zero modes, which would correspond to the world-sheet 
fermions of the fundamental string. Instead, the object that is S-dual
to the fundamental string corresponds to a combination of both types
of D-strings.\footnote{Such a combination has been discussed in the
context of D3-branes in \cite{KT}, where it was termed a `self-dual'
3-brane.} The fermions are then provided by open strings which begin
on one D-string and end on the other. It follows from (\ref{D0}) that
the combination is only charged under the untwisted R-R fields, and
therefore that S-duality exchanges the NS-NS 2-form with the R-R
2-form in the {\em untwisted} sector. The twisted fields are invariant
under the duality.  

Such a combination should appear as a {\em bound state} of the two
D-strings. In the following section we shall argue that an analogous
bound state of the two different Type 0A D-particles forms in the
regime where the tachyon becomes massive. This suggests that the same
is true for the two different Type 0B D-strings, namely that they form
a bound state in nine-dimensions when $g^{(0B)}/R^{(0B)}$ is large 
enough to make the tachyon massive. The ten-dimensional picture is
somewhat less clear.

\section{Type 0A fermions}
\setcounter{equation}{0} 

In the M-theory orbifold the fermions carry odd Kaluza-Klein momentum,
and therefore have a minimal (classical) mass  
\be
\label{fermmass}
 m_f = {1\over R^{(M)}} \,.
\ee
In the limit $R^{(M)}\rightarrow 0$ the fermions decouple, and we
obtain a purely bosonic spectrum. However, as in the case of ordinary
Kaluza-Klein excitations of M-theory on ${\bf S}^1$, the existence of
states with mass (\ref{fermmass}) suggests that Type 0A has
corresponding non-perturbative (Dirichlet-brane-like) states whose
mass is proportional to $1/g^{(0A)}$. In particular,
eq.~(\ref{fermmass}) implies that Type 0A (which perturbatively is a
purely bosonic theory) contains fermionic solitons.

Strictly speaking, the above expression for the mass can only be
trusted in the classical limit, {\it i.e.} for 
$R^{(M)}\gg M_{11}^{-1}$. Due to the lack of supersymmetry,
masses are not protected against quantum corrections at small
$R^{(M)}$. In fact, since there is a tachyon in the spectrum, one
expects that the masses are renormalized by an infinite amount. As a
consequence, one should not expect to find any fermions in weakly
coupled Type 0A string theory; the fermions should only appear when
the coupling constant is sufficiently large to remove the tachyon.

However as we shall now explain, all the ingredients that are
necessary to form these fermions are already present in weakly coupled
Type 0A theory. Recall that the theory has two kinds of D-particles,
$\ket{D0,+}$ and $\ket{D0,-}$. Since open strings beginning and ending
on the same D-particle are purely bosonic, each of the D-particles is
bosonic, and has a non-degenerate ground state. Strings beginning on
$\ket{D0,+}$ and ending on $\ket{D0,-}$, on the other hand, are purely
fermionic. A configuration of two coincident D-particles of different
type would therefore possess fermionic  zero-modes, whose quantization 
would result in a supermultiplet ground state, and thus spacetime
fermions. More realistically, the two D-particles might form a bound
state with a non-vanishing separation. The fermionic modes would then
be massive, and this would lead to a non-degenerate bosonic ground
state and at least one stable fermionic excited state. 

We shall now argue that such a bound state can form when the tachyon 
becomes massive. To lowest order, the potential between unlike
D-particles is given in the open string (loop) channel by
\be
\label{loop1}
V_{+-}(r) =  2 \int {dt\over 2t} \int {dk^0\over 2\pi}
               \Tr_{R}[e^{-2\pi\alpha' t(k^2+M^2(r))} 
               {1\over 2}(1+(-1)^F)] \,,
\ee
where $r$ is the separation between the D-particles. Upon integration
of $k^0$, we find
\be
V_{+-}(r) = \int{dt\over 2t} (8\pi^2\alpha' t)^{-1/2}
                e^{-tr^2/(2\pi\alpha')}
                {f_2^8(q) \over f_1^8(q)} \,,
\ee
where $f_i$ are the standard $f$ functions \cite{PolCai}, and 
$q=\exp(- 2 \pi t)$. At short distances, $r\ll\sqrt{\alpha'}$, we can 
approximate $V_{+-}$ by taking the limit $t\rightarrow \infty$, 
in which case
\be
  {f_2^8(q) \over f_1^8(q)} \sim  16 + O(e^{-\pi t})\,.
\ee
In this approximation, the potential becomes 
\footnote{The expression for the potential is similar to the D0-D8
case in Type IIA \cite{D0D8}, and is evaluated using the analytic 
continuation of the gamma function to negative argument.}
\be
   V^{short}_{+-}(r) \sim  {8  \over \sqrt{8 \pi^2 \alpha'}}
   \int {dt\over t^{3/2}} e^{-tr^2 / (2\pi\alpha')}
  = -{4\over\pi\alpha'}|r|\,.
\ee
In contrast, the potential between like D-particles is given by 
the open string amplitude
\begin{eqnarray}
\label{loop2}
V_{++}(r) &=& \mbox{} - 2 \int {dt\over 2t} \int {dk^0\over 2\pi}
               \Tr_{NS}[e^{-2\pi\alpha' t(k^2+M^2(r))} 
               {1\over 2}(1+(-1)^F)] \nonumber \\
 & =& \mbox{} - \int{dt\over 2t} (8\pi^2\alpha' t)^{-1/2}
                e^{-tr^2/(2\pi\alpha')}
                {f_3^8(q) - f_4^8(q) \over f_1^8(q)} \,,
\end{eqnarray}
which at short distances becomes
\be
   V^{short}_{++}(r) \sim {4\over\pi\alpha'}|r|\,.
\ee
We therefore find that the potential between like D-particles is
attractive at short distances, whereas the potential between unlike
D-particles is repulsive.

At long distances the picture is complicated by the tachyon.
For $r\gg\sqrt{\alpha'}$ we can approximate the amplitude by taking
the limit  $t\rightarrow 0$, where
\be
 {f_2^8 (q) \over f_1^8 (q)} = {f_3^8(q) - f_4^8(q)\over f_1^8(q)}
  \sim t^4\Big(e^{\pi/t} - 8 + O(e^{-\pi/t})
  \Big)\,.
\ee
The first dominant term corresponds to the exchange of the closed string
tachyon, and gives a divergent contribution to the potential. It
is attractive for like D-particles and repulsive for unlike ones.
The second term corresponds to the exchange of massless fields,
and gives a {\em repulsive} contribution for {\em like} D-particles and an
{\em attractive} one for {\em unlike} D-particles. Indeed, like D-particles
interact through massless R-R fields as well as through massless NS-NS
fields (gravity and dilaton). The former is repulsive, and twice as
strong as the latter, which is attractive. The net interaction due to
massless fields is therefore repulsive. Unlike D-particles, on the
other hand, are charged under different R-R fields, and so interact
only through the massless NS-NS fields; their interaction is therefore
attractive at the massless level. 

As we have learned in the previous section, at strong coupling the
Type 0A tachyon acquires a mass, whereas the massless NS-NS fields
remain massless. Although we cannot directly trust either the short or
the long distance calculation of the potential at strong coupling, the
long distance force will be dominated by the exchange of the
lowest-lying closed string states, and this implies, whatever the
details, that the long-distance potential between two different
D0-branes is  {\em attractive} at strong Type 0A coupling. Thus at
strong 0A coupling, the bound state is likely to exist, and it has the
right properties to correspond to the fermionic KK excitations of the
M-theory orbifold.

\section{Non-perturbative consistency of open string models}
\setcounter{equation}{0} 

The Type 0 open string models of table~1 were constructed by
projecting Type 0A and Type 0B string theory either by the world-sheet
parity operator $\Omega$, or by a combination of $\Omega$ with
another $\Zop_2$ symmetry. Given our conjecture about the
eleven-dimensional origin of Type 0A and Type 0B, it is worthwhile to
re-examine these models in the context of M-theory.

It is well known that $\Omega$ lifts to the operation that inverts
the sign of the 3-form. This by itself is not a symmetry of M-theory
as it reverses the orientation of the membrane. The Type 0A open
string model can therefore not be related to M-theory, and is
thus inconsistent at non-zero coupling. 
Another way of reaching the same conclusion is to observe that (at
zero coupling) $\Omega$ exchanges the massless R-R fields in the
untwisted sector of Type 0A with those in the twisted sector. As the
coupling is increased the twisted fields become massive, and $\Omega$
does not define a symmetry any more. This again implies that the
corresponding open string model is only consistent at zero coupling.

The consistency of the Type 0B models is most easily analysed by
considering their T-dual Type 0A models after compactification on a
circle. There are three cases to consider. The 
orientifold of Type 0B by $\Omega$ is related by T-duality to Type 0A
on ${\bf S^1}/\Omega\cdot I$,
where $I$ denotes a reflection in the compact direction.
In terms of M-theory this projection
corresponds to the symmetry (\ref{reflect}), and therefore the model
is presumably consistent for all values of the coupling constant. 

The T-dual of the second orientifold model of Type 0B by
$\widehat\Omega=\Omega (-1)^{F^s_R}$ is Type 0A on 
${\bf S^1}/\Omega (-1)^{F^s_R} \cdot I$. This differs from the first 
orientifold projection by the operator $(-1)^{F^s_R}$, that acts as 
$+1 (-1)$ on all states in the untwisted  NS-NS (R-R) sector
of Type 0A. This does not lift to a symmetry of M-theory, and
therefore the associated orientifold model is inconsistent at finite
coupling. 

Finally, the third Type 0B model, involving $\Omega'=\Omega(-1)^{F_R}$,
is related by T-duality to Type 0A on 
${\bf S^1}/\Omega (-1)^{F_R} \cdot I$. This differs from the first
orientifold projection by $(-1)^{F_R}$, that acts as $+1$ $(-1)$   
on the untwisted (twisted) sector of the M-theory orbifold. This
presumably defines a symmetry of M-theory, but since we do not have
detailed knowledge of the twisted sector, we cannot check this
further.

\section{Conclusions}
\setcounter{equation}{0}

In this paper we have argued that two of the non-supersymmetric
closed string theories in ten dimensions, Type 0A and Type 0B, fit
into the framework of M-theory. We have shown that the spectrum of
Type 0A agrees, in a certain limit, with that of the orbifold of
M-theory on {\bf S}$^1/(-1)^{F^s}\cdot S$. We have also explained how
the fermionic states of the M-theory orbifold might be accounted for
in terms of D-particles in Type 0A. One consequence of this conjecture
is that the tachyon of Type 0A becomes massive at strong coupling.  
The conjecture also implies that Type 0B is self-dual, and that some
of the previously obtained Type 0 open string models are inconsistent
at finite coupling.  

A crucial role in our discussion was played by the nine-dimensional
theory that interpolates between Type IIA at infinite radius and
Type 0A at zero radius. This theory has a non-vanishing
one-loop cosmological constant which diverges at the radius where the
tachyon appears \cite{Rohm}. The size of the circle is therefore not a
modulus of the theory, and there exists a non-trivial potential which
is minimized at infinite radius; for sufficiently large radii
the theory therefore decompactifies. This suggests that the same is
true for the ten-dimensional interpolating theory, and therefore that
Type 0A will decompactify dynamically to eleven dimension at
sufficiently strong coupling. 

It would be interesting to understand how the other non-supersymmetric
theories fit into the M-theory framework. Some time ago we argued that
one of the orientifolds of Type 0B is dual to the compactification of
the 26-dimensional bosonic string on the $SO(32)$ weight lattice
\cite{BG1}, but the relations of the other non-supersymmetric theories
to one another (and to M-theory) remain to be understood. 
Some progress in this direction has also been made in
\cite{BD}, but much still remains to be uncovered.

\section*{Acknowledgements}

We would like to thank Igor Klebanov for very useful conversations,
and Ralph Blumenhagen and Shigeki Sugimoto for pointing out 
a couple of errors in an earlier version of this paper.
O.B. is supported in part by the DOE under grant no. DE-FG03-92-ER
40701. M.R.G. is supported by a College Lectureship of Fitzwilliam
College, Cambridge.

\end{document}